\documentclass{pnastwo}
\usepackage[dvips]{graphicx}
\usepackage{amssymb,amsfonts,amsmath}
\def \bea{\begin{eqnarray}}
\def \eea{\end{eqnarray}}

\begin{document}

\title{Order-by-disorder in the antiferromagnetic Ising model on an elastic triangular lattice}

\author{Yair Shokef\affil{1}{Department of Materials and Interfaces, Weizmann Institute of Science, Rehovot 76100, Israel}, Anton Souslov\affil{2}{Department of Physics and Astronomy, University of Pennsylvania, Philadelphia, PA 19104, USA} \and Tom C. Lubensky\affil{2}{}}

\maketitle

\begin{article}

\begin{abstract}
Geometrically frustrated materials have a ground-state degeneracy that
may be lifted by subtle effects, such as higher order interactions
causing small energetic preferences for ordered structures.
Alternatively, ordering may result from entropic differences between
configurations in an effect termed order-by-disorder. Motivated by
recent experiments in a frustrated colloidal system in which ordering
is suspected to result from entropy, we consider in this paper the
antiferromagnetic Ising model on a deformable triangular lattice. We
calculate the displacements exactly at the microscopic level, and
contrary to previous studies, find a partially disordered ground state
of randomly zigzagging stripes. Each such configuration is deformed
differently and thus has a unique phonon spectrum with distinct
entropy, lifting the degeneracy at finite temperature. Nonetheless, due
to the free-energy barriers between the ground-state configurations,
the system falls into a disordered glassy state.
\end{abstract}


\abbreviations{FCC, face-centered-cubic; HCP, hexagonal-close-packing}

\dropcap{F}rustrated systems are characterized by interactions that may
not be satisfied simultaneously~\cite{PhysicsToday}. This leads to a
degenerate and thus disordered ground state, and naively one would
expect to have disorder down to zero temperature~\cite{Wannier}.
However, the frustrated phase is very sensitive to small perturbations
that can order it. These include anisotropic~\cite{Houtappel} or
longer-range~\cite{Metcalf} interactions and lattice
deformations~\cite{Kardar,Bulbul,Terao,Yamashita,Tchernyshyov,Schilling}
as well as entropic effects that may lift the ground-state degeneracy
at finite temperature, in a process termed
order-by-disorder~\cite{Villain,Henley87,Henley89,Chubukov,Reimers,Bergman}.
Frustration and its relief due to order-by-disorder are traditionally
investigated in antiferromagnets and, in particular, in compounds that
have a triangular lattice
structure~\cite{Collins,Maignan,WangVishwanath,Starykh}. Recent
experiments have demonstrated that artificial systems made of
mesoscopic building blocks such as single-domain magnetic
islands~\cite{Wang} or colloidal spheres~\cite{Han} exhibit behavior
which is similar to that of magnetic systems comprised of atomic-scale
particles. Such mesoscopic systems enable direct visualization of the
dynamics at the single-particle level and thus provide insight into the
microscopic physical mechanisms responsible for the peculiar properties
of frustrated matter. In this paper we address the theoretical
underpinnings of these new experimental systems and thus provide a
vital step in understanding their unusual behavior and its connection
to atomic-scale antiferromagnets.

A densely-packed monolayer of hard spheres buckles out of its plane
when it is confined between walls that are separated by slightly more
than a single sphere diameter~\cite{Chou}. Entropic forces depending
only on geometry give rise to effective anti-ferromagnetic interactions
favoring motion of neighboring spheres toward opposite
walls~\cite{Shokef}, leading at high densities to stripes of
alternating up and down spheres. The close-packed state is highly
degenerate: the same maximal density is obtained by straight stripes or
by any set of parallel stripes that zigzag within the
hexagonally-packed
layer~\cite{Han,Shokef,Schmidt96,Schmidt97,Zangi98,Melby}. Recent
experiments indicate a possible preference of the stripes to be
straight rather than to zigzag randomly in the plane, suggesting that
at densities below close-packing there is an order-by-disorder effect
giving an entropic advantage for straight stripes~\cite{Han}.

This quasi-two-dimensional problem is strongly related to the old, yet
unsolved question of what is the stable high-density structure of hard
spheres, face-centered-cubic (FCC) or hexagonal-close-packing
(HCP)~\cite{Stillinger,Alder,Rudd}. In three dimensions, maximal
density is obtained by stacking hexagonally packed layers with
arbitrary sideways shifts between the close-packed positions. And, as
for the buckled monolayer, slightly below close-packing it is not clear
which structure has the greatest entropy. Experiments on colloidal
crystals that were grown slowly enough exhibit FCC order~\cite{Pusey},
while there is controversy on whether the theoretical estimates have
reached the accuracy required to resolve the elusive entropic
difference between FCC and
HCP~\cite{Woodcock1,Bolhuis,Woodcock2,Mau,Radin,Koch}.

As for the FCC \emph{vs} HCP question, the entropy, or free
volume, of buckled hard spheres is a collective function of the
positions of all particles, thus the ability to obtain analytical
results for it is very limited. Thus, instead of {\em approximating}
the entropy of this hard-sphere system, we consider the
anti-ferromagentic Ising model on a deformable triangular lattice.
This model has the same degenerate ground state of zigzagging
stripes as the colloidal system, and for it we can {\em
exactly} calculate the free-energy difference between the competing
configurations. We find that straight stripes are always
favored entropically. However, the free-energy barriers between various
ground states are huge compared to this entropic gain, causing the
system to fall into a disordered glassy state upon cooling.

\begin{figure}
\centerline{\includegraphics[width=0.5\textwidth]{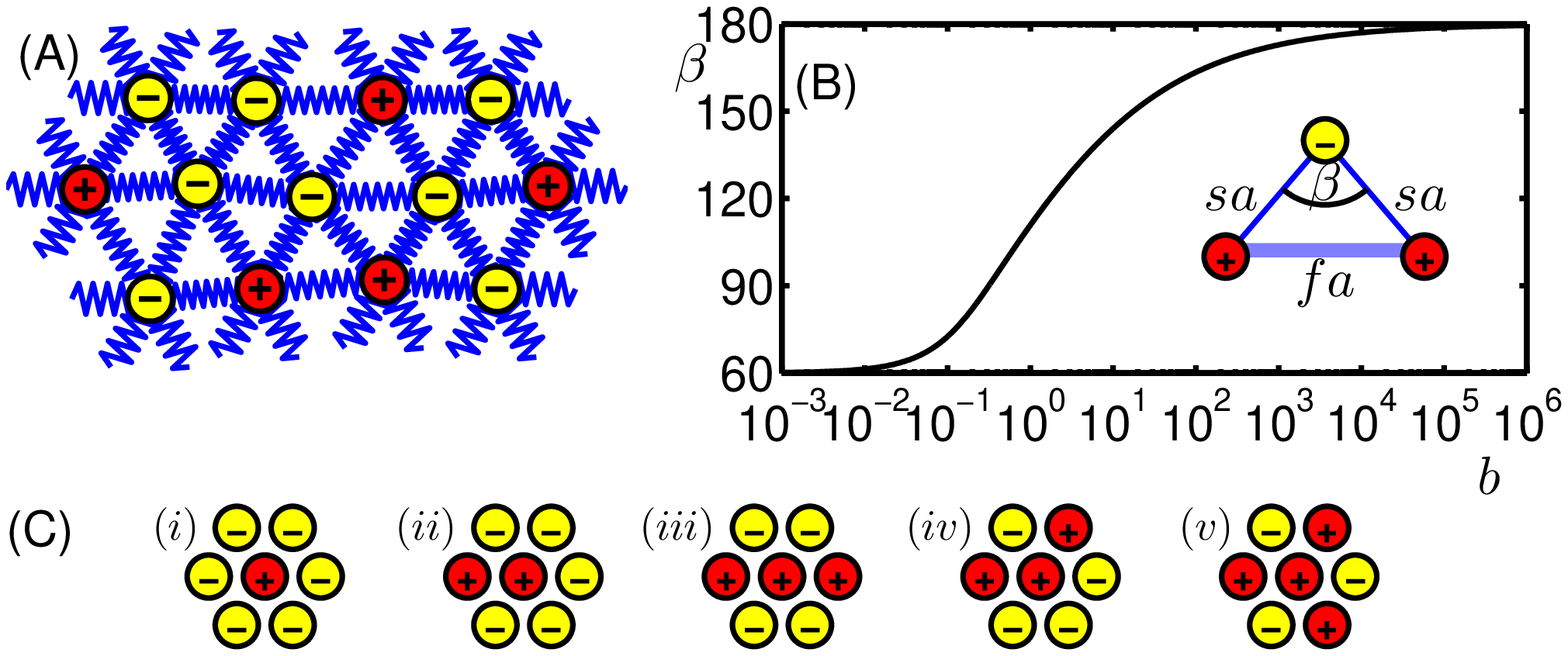}}
\caption{Model and its ground state: A) Two-dimensional triangular
network of antiferromagnetic Ising spins connected with harmonic
springs. B) Deformation angle $\beta$ vs the dimensionless ratio $b
\equiv \frac{J\epsilon}{Ka}$ between the magnetoelastic interaction
strength and the lattice rigidity, from Eq.~(\ref{eq:sf}). Inset:
Isosceles triangle with head angle $\beta$, satisfied bonds (thin
lines) compressed by $s$ and frustrated bond (thick line) stretched by
$f$. C) The five possible configurations (up to rotations and spin
inversions) of a particle and its neighbors, with a single frustrated
bond in each triangle.}
\label{fig:model}
\end{figure}

In our Ising model, each site $i$ at continuous position $\vec{r}_i$ on
a triangular network is occupied by a particle of discrete spin
$\sigma_i = \pm 1$, and each nearest neighbor bond comprises a
harmonic spring of stiffness $K$ and relaxed length $a$, see
Fig.~\ref{fig:model}A. The internal energy of the system depends only
on the spin product $\sigma_i \sigma_j$ of neighboring particles and on
their relative positions, $\delta r_{ij} = |\vec{r}_i - \vec{r}_j| -
a$. The model's Hamiltonian is given by the following sum over all
nearest-neighbor pairs $\langle i j \rangle$,
\bea 
\mathcal{H} =
\sum_{\langle i j \rangle} \bigg[ \left(1 - \epsilon \delta r_{ij}
\right) J \sigma_i \sigma_j + \frac{K}{2} \delta r_{ij} ^2 \bigg] .
\label{eq:potential_energy}
\eea
The antiferromagnetic interaction
strength is equal to $J>0$ for particles separated by the relaxed
spring length $a$, and decreases linearly with distance at a rate
$\epsilon>0$. 
This mimics the buckled colloidal monolayer in the following way: For
hard spheres, free energy is determined by their free volume. However,
it is convenient to reduce this to an effective two-particle repulsive
interaction. While the particles are confined to be
quasi-two-dimensional, they do have a limited freedom to move in the
vertical direction, and thus nearest neighbors prefer to sit at
opposite heights. We can now map the vertical position of each
particle onto an Ising degree of freedom, due to a preference for
either an ``up'' or ``down'' state. This leads to an effective
two-dimensional antiferromagnetic Ising model, with the coupling between the
elastic and the magnetic degrees of freedom as reflected in our
Hamiltonian~(\ref{eq:potential_energy}), i.e. the effective antiferromagnetic interaction decays as
the in-plane separation increases~\cite{Shokef}.

We model the in-plane entropic repulsion between neighboring spheres by
a rotationally invariant central-force harmonic spring
[Eq.~(\ref{eq:potential_energy})]. Approximate versions of this model
have been studied previously: Ref.~\cite{Bulbul} uses a
linearized long-wavelength elastic energy, invariant with respect to
only infinitesimal rotations, that limits deformations to be small and
to vary slowly in space, while in Ref.~\cite{Kardar}
deformations are taken to be uniform.

In this paper we present results for the case in which the total area
of the system is fixed to the area of a triangular lattice with lattice
constant equal to $a$. However, we obtained qualitatively
similar results for systems compressed or dilated with respect to this
simple situation. From the equivalence between ensembles we
thus expect to get the same results also when considering the fixed
pressure case, which is probably more appropriate for the colloidal
experiments.

As in the rigid triangular-lattice model~\cite{Wannier,Houtappel}, the
antiferromagnetic interactions along each three-particle loop in our
deformable network cannot be satisfied simultaneously, and energy is
minimized by having two satisfied ($\sigma_i \sigma_j = -1$) bonds and
a single frustrated ($\sigma_i \sigma_j = 1$) bond around each triangular
plaquette. Because of the magnetoelastic coupling, energy may be
lowered by stretching the frustrated bonds by a factor $f$ and
compressing the satisfied ones by a factor $s$. We fix the area of each
resulting isosceles triangle to be that of the initial equilateral
plaquette with sides $a$, and thus we can parametrize the deformation
by the head angle $\beta$ (see Fig.~\ref{fig:model}B, inset): $f(\beta)
= 3^{1/4} (\tan \frac{\beta}{2} )^{1/2}$, $s(\beta) = 3^{1/4} \left(2
\sin \beta \right)^{-1/2}$. Minimizing the
energy~(\ref{eq:potential_energy}) with respect to $\beta$ yields (see SI Appendix)
\bea
(2s'-f')J\epsilon+\left[2(s-1)s'+(f-1)f'\right]Ka=0 . \label{eq:sf} 
\eea
Figure~\ref{fig:model}B shows how the triangles deform from
$\beta=60^\circ$ toward $\beta=180^\circ$ as the ratio $b\equiv
\frac{J\epsilon}{Ka}$ of the magnetoelastic interaction strength to the
lattice rigidity grows.

\begin{figure}
\centerline{\includegraphics[width=0.5\textwidth]{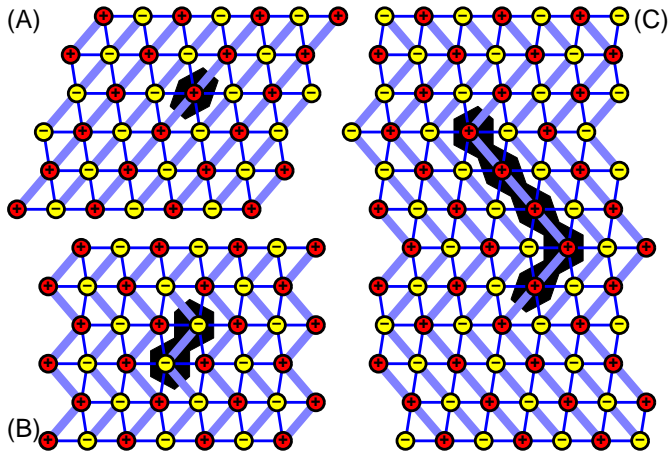}}
\caption{Zigzagging stripes: Ground-state configurations are generated
by stacking layers of particles with alternating spin and with
arbitrary relative polarities between successive layers. The figure
shows only simple configurations with periodic sequences of the
straight $(iii)$ and bent $(iv)$ segments, as defined in
Fig.~\ref{fig:model}C. However ground-state configurations do not
necessarily have a finite unit cell. A) Straight stripes, for which all
particles are in the state $(iii)$. B) Bent stripes, for which all
particles are in state $(iv)$. C) Zigzagging stripes with a more
complicated unit cell comprised of both $(iii)$ and $(iv)$. The shaded
regions represent the unit cells used for the low-temperature expansion
explained in the text. (A), (B), and (C) have one, two and five
particles per unit cell, respectively.
Thick blue lines represents the stretched frustrated bonds and thin blue lines the compressed bonds for which the antiferromagnetic interaction is satisfied.}
\label{fig:ZZ}
\end{figure}

Thus, each plaquette of the triangular lattice would minimize its
energy by deforming into an isosceles triangle. We now show how this
can be accommodated in the ground state by global deformations of the
system. In the rigid triangular lattice, since each triangle must have
exactly one frustrated bond, five nearest-neighbor configurations are
allowed in the ground state (Fig.\ref{fig:model}C)~\cite{Ogawa}.
Requiring that the angle opposing each frustrated bond deforms to
$\beta > 60^\circ$ selects configurations $(iii)$ and
$(iv)$~\cite{Han,Shokef}, which give rise to zigzagging
stripes. Thus, selecting a stripe of frustrated bonds defines a ground
state of the lattice. It can be constructed by starting with a row of
alternating spins (along the horizontal axis) and stacking copies of
this row (along the vertical axis), as shown in Fig.~\ref{fig:ZZ}. The
intra-row and inter-row separations are $sa$ and $s a \sin \beta$,
respectively, and the lateral shifts are determined by the arbitrary
polarity of each row with respect to the one preceding it. Since these
rows of alternating spins may be in any of the three principal
directions of the network, and for a system of $N$ particles, each of
the $\sqrt{N}$ rows may be in one of two states, the ground-state
degeneracy is $3 \cdot 2^{\sqrt{N}}$.

This partially-disordered ground state is realized by deformations that
may vary rapidly in space; if deformations are assumed to be
homogeneous~\cite{Kardar} or to vary slowly~\cite{Bulbul}, straight
stripes are selected. In particular, the zigzagging stripes
that minimize our microscopic Hamiltonian (\ref{eq:potential_energy})
have a higher energy than straight stripes in the
coarse-grained Hamiltonian considered in~\cite{Bulbul}. The ground
state of zigzagging stripes that we find here is precisely the state
that maximizes the packing density of buckled
spheres~\cite{Han,Shokef}, which constitutes the connection of our
model to the colloidal system.

Given the high ground-state degeneracy and the fact that it takes a
discrete energy (of order $J$) to flip a spin in the ground state, one
might naively expect to find a stable phase of randomly zigzagging
stripes at sufficiently low temperature ($k_BT \ll J$). However, at positive
temperature, the entropy of particle fluctuations around the
energy-minimizing position is different in the different ground-state
configurations. We will show that the state with straight stripes has
lower free energy (or greater entropy) than other states, making it the
stable thermodynamic phase at arbitrarily low temperature. Nonetheless,
the resulting free-energy differences between the different zigzagging
realizations are much smaller than the free-energy barriers between the
ground-state configurations and the system typically falls into
disordered meta-stable configurations.

\begin{figure}
\centerline{\includegraphics[width=0.5\textwidth]{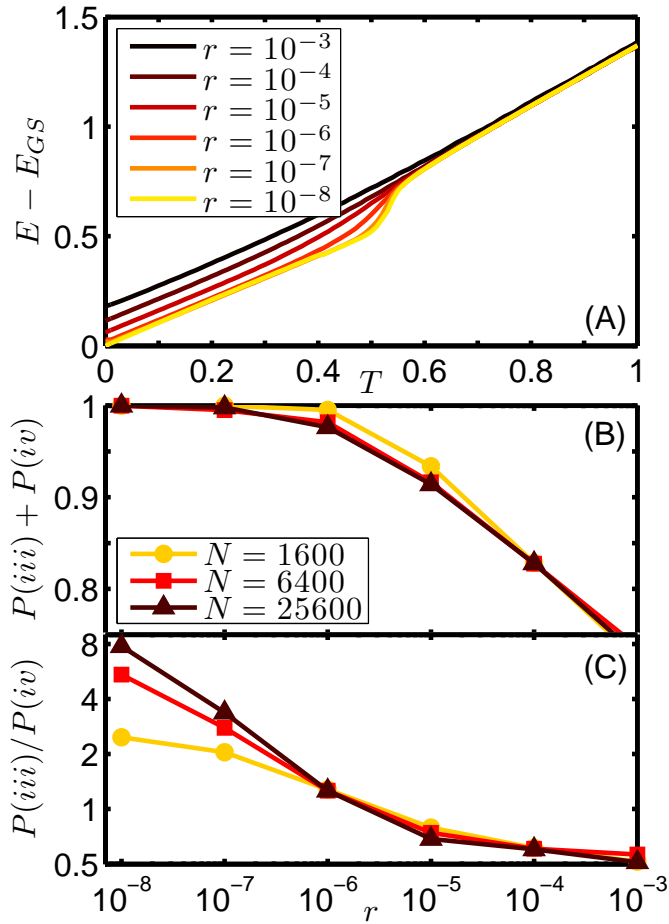}}
\caption{Monte-Carlo simulations: A) Energy $E$ per particle above the
ground-state energy $E_{GS}$ vs temperature $T$ for different cooling
rates $r$ as indicated in the legend. System size is $N=6400$. Similar
results were obtained for $N=1600$ and $N=25600$. B,C) Probabilities of
finding the local configurations $(iii)$ and $(iv)$ defined in
Fig.~\ref{fig:model}C at $T=0$ following cooling at different rates $r$
for various system sizes $N$, as indicated in the legend. Note the
logarithmic scale for the ratio $P(iii)/P(iv)$ in (C). Error bars are
smaller than the symbols. Model parameters in all simulations are
$J=1$, $\epsilon=2$, $K=8$, $a=1$. 
Thus $b \equiv \frac{J\epsilon}{Ka}=0.25$. This yields 
a deformation angle of
$\beta=86^\circ$ in the ground state and a difference in the
free-energy coefficient, Eq.~(\ref{eq:A}), of $d\mathcal{A}0.05$ between
straight and bent stripes.}
\label{fig:MC}
\end{figure}

To demonstrate the peculiar slow dynamics of this model, we used
Monte-Carlo simulations in which individual particles can flip their
spin or move continuously in the plane and in which the simulation box
may change its shape to accommodate the global deformations of the
network~\cite{Shokef}. We start from a disordered state at high
temperature, and we slowly cool the system at a rate $r$ (time is
measured in attempted Monte-Carlo steps per particle).
Figure~\ref{fig:MC}A shows that at high temperature the system follows
an equilibrium curve irrespective of cooling rate, while below a
certain temperature ($T \approx 0.6$ for the parameter values shown
here), the system's energy has a clear cooling rate dependence.
Apparently, if the cooling is slow enough ($r \le 10^{-6}$ here), the
system manages to reach the ground state.

\begin{figure}
\centerline{\includegraphics[width=0.5\textwidth]{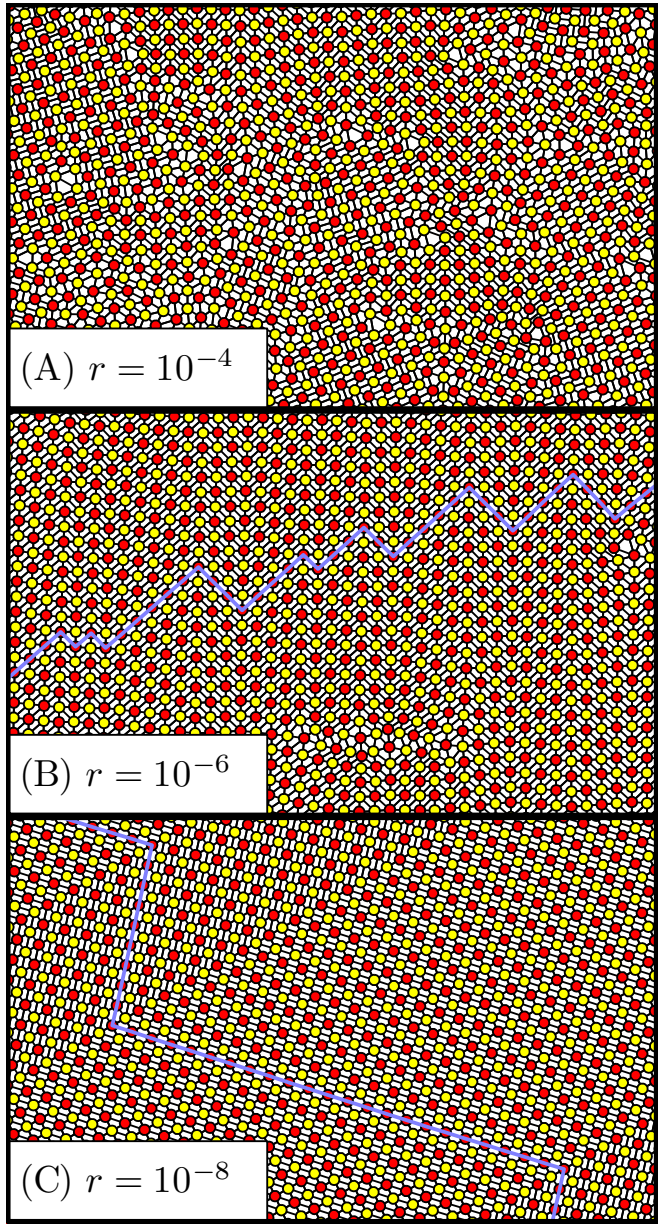}}
\caption{Ordering with decreasing cooling rate: Portion of the system
in its final configuration after cooling at rates $r=10^{-4}$(A),
$10^{-6}$(B), $10^{-8}$(C). Model parameters are $J=1$, $\epsilon=2$,
$K=8$, $a=1$. 
Thus, $b \equiv \frac{J\epsilon}{Ka}=0.25$, and in the ground state each triangular plaquette is deformed to an isosceles with head angle $\beta=86^\circ$.
System size is $N=25600$. For the fastest cooling rate,
the system falls into a disordered state (A), for which the fraction of
particles in the local configuration of a straight stripe (see
Fig.~\ref{fig:model}C) is $P(iii)=0.31$ and the fraction in that of
bends is $P(iv)=0.52$. For the slower cooling rates, zigzagging stripes
are formed (B,C). The blue lines are guides to the eye, which highlight
a line of frustrated bonds, and emphasize that as the cooling rate is
decreased from (B) to (C), the stripes become more straight. In (B),
$P(iii)=0.54$ and $P(iv)=0.44$, whereas in (C), $P(iii)=0.9$ and
$P(iv)=0.1$.}
\label{fig:xyz}
\end{figure}

Figure~\ref{fig:xyz}A shows that indeed if the system is cooled too
rapidly ($r=10^{-4}$) it falls into a disordered state with multiple
small domains of a local stripy structure. For a slower cooling rate
($r=10^{-6}$, Fig.~\ref{fig:xyz}B), the system finds a ground-state
configuration with zigzagging stripes, such that the local
configurations $(iii)$ and $(iv)$ defined in Fig.~\ref{fig:model}C are
roughly equally represented, thus corroborating this phase as randomly
zigzagging stripes. Figure~\ref{fig:xyz}C shows that for even slower
cooling ($r=10^{-8}$), there is preference for ground-state
configurations in which the stripes are more straight than bent, namely
configuration $(iii)$ is preferred over $(iv)$. This is quantified in
Fig.~\ref{fig:MC}, B, C, which shows results obtained by averaging over
multiple realizations for each system size and cooling rate. For very
fast cooling, the system remains quite disordered with $P(iii)/P(iv)
\approx 0.5$, in accordance with the combinatoric weights of these two
local configurations. For $r \le 10^{-6}$, the system manages to find
its ground state, as seen by the fact that $P(iii)+P(iv)=1$, namely all
particles are in one of these two states. For $r=10^{-6}$ these two
states are equally probable, $P(iii) \approx P(iv)$, as one would
expect for randomly zigzagging stripes. However, as the cooling rate is
decreased even more, there is a clear preference for the straight
configuration $(iii)$ over the bent one $(iv)$. This is in qualitative
agreement with recent experiments~\cite{jen} exploring the cooling rate
dependence of our colloidal antiferromagnet~\cite{Han}.

Although the Ising degrees of freedom in our model do not have
natural physical dynamics associated with them, the Monte-Carlo
dynamics we employ are useful in exploring actual out-of-equilibrium
dynamics of spin systems~\cite{Tome,Sides,Korniss,Robb}. Thus, to our understanding, the
cooling rate in our simulations should be interpreted as being directly
proportional to the experimental cooling rate, and it
would be interesting to explore the connection between the transition rates we
apply in our Monte-Carlo scheme and the actual physical quantities in
the experiments.

The preference we observe for straight over bent stripes comes from an entropic
difference between the two, as is explained in the following
low-temperature expansion. Expanding the Hamiltonian about the
fixed-spin ground state to lowest order, we find terms quadratic in
particle displacements. Thus, the lowest-energy excitations are
harmonic modes of vibrations with frequencies which we denote by
$\{\omega_{k}\}$ (note that $k$ indexes the normal modes and does not
necessarily refer to a wavevector). We assume the temperature is low
enough ($k_B T \ll J$) to ignore spin flips, yet high enough ($k_B T
\gg \hbar \omega$) to ignore quantum effects. Therefore, we use the
canonical partition function of a classical harmonic oscillator of
frequency $\omega$, $Z \propto k_BT / \omega$, to write the free energy
of the system as the following sum over all normal modes 
\bea 
\mathcal{F} = -k_B T \log Z = -k_B T \left[ \sum_k{\log\left(\frac{k_BT}{
\omega_k}\right)} + {\rm const.} \right] \label{eq:FFF} 
\eea
We emphasize at this point that the frequencies $\{\omega_{k}\}$ refer to the fast oscillations of the positional degrees of freedom, which at low enough temperature are expected to be much faster than the spin flips.

The different spin states $\{\sigma_i\}$ of the various ground-state
configurations impose different deformations $\{\vec{r}_i\}$, thus each
one has a distinct spectrum $\{\omega_k\}$ of eigen-frequencies, and a
different entropy, which is expressed as a different temperature
dependence of the free energy $\mathcal{F}$ in Eq.~(\ref{eq:FFF}). At $T=0$,
entropy is irrelevant and all ground-state configurations are
equivalent, but at $T>0$ this degeneracy is lifted and the
configuration with the minimal free energy is selected. The frequencies
of all vibrational modes scale as $\omega_k^2 \propto K/M$, where $M$
is the mass of each particle. The normalized spectrum depends on the
spin state $\{\sigma_i\}$ and on the deformation angle $\beta$, which
is set by $b \equiv \frac{J\epsilon}{Ka}$ (see Eq.~(\ref{eq:sf}) and
Fig.~\ref{fig:model}B). Hence we define
\bea 
\mathcal{A} (\{\sigma_i\},\beta) \equiv \frac{1}{N} \sum_k\log\left( \sqrt{\frac{M}{K}} \omega_k\right)
, \label{eq:A}
\eea
and write $\mathcal{F} = Nk_BT \left[ \mathcal{A} (\{ \sigma_i \} ,
\beta) + {\rm const.} \right]$. Thus minimizing $\mathcal{F}$ is equivalent to
minimizing $\mathcal{A}$.

\begin{figure}
\centerline{\includegraphics[width=0.5\textwidth]{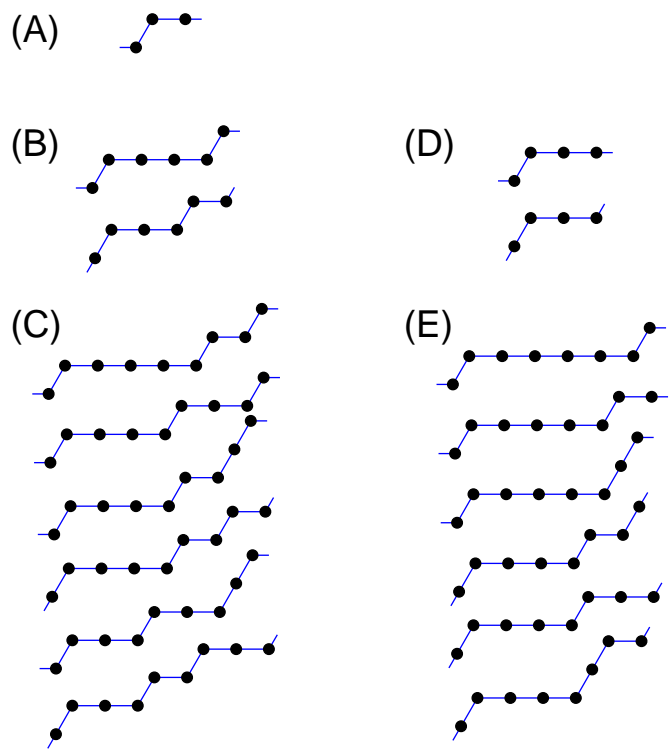}}
\caption{Examples for some of the unit cells used in evaluating
the free energy in Fig.~\ref{fig:free_energy}: All unit cells with
$p_s=1/3$ containing 3 (A), 6 (B), and 9 (C) particles, and with
$p_s=1/2$ and 4 (D) and 8 (E) particles.}
\label{fig:samp_unit_cell}
\end{figure}

We can analytically calculate the normal modes of vibrations
of the deformed lattice for any ground state with a periodic repetition
of straight $(iii)$ and bent $(iv)$ segments (see SI Appendix). In such cases the normal
modes are phonons and the index $k$ labeling them may be associated
with their wavevector. The free-energy coefficient, $\mathcal{A}$, may then be
obtained by numerically summing over the appropriate Brillouin zone.
Figure~\ref{fig:ZZ} highlights the unit cells in ground states with
periods of one, two, and five particles. Now we consider ground
states with larger unit cells, and show that the free energy is
mainly
determined by the fraction $p_s \equiv P(iii)/[P(iii)+P(iv)]$ of
straight segments. To clarify which configurations we used in these
calculations, we show in Fig.~\ref{fig:samp_unit_cell} several
representative unit cells. For example, $p_s=1/3$ in a single
configuration with three particles in its unit cell
(Fig.~\ref{fig:samp_unit_cell}A), but also in two configurations with
six particles (Fig.~\ref{fig:samp_unit_cell}B), in six with nine
particle (Fig.~\ref{fig:samp_unit_cell}C), and in an increasing number
of other configurations with larger unit cells. Similarly, for
$p_s=1/2$, we show the two configurations with four particles per unit
cell in Fig.~\ref{fig:samp_unit_cell}D and the six with eight
particles in Fig.~\ref{fig:samp_unit_cell}E.

\begin{figure}
\centerline{\includegraphics[width=0.5\textwidth]{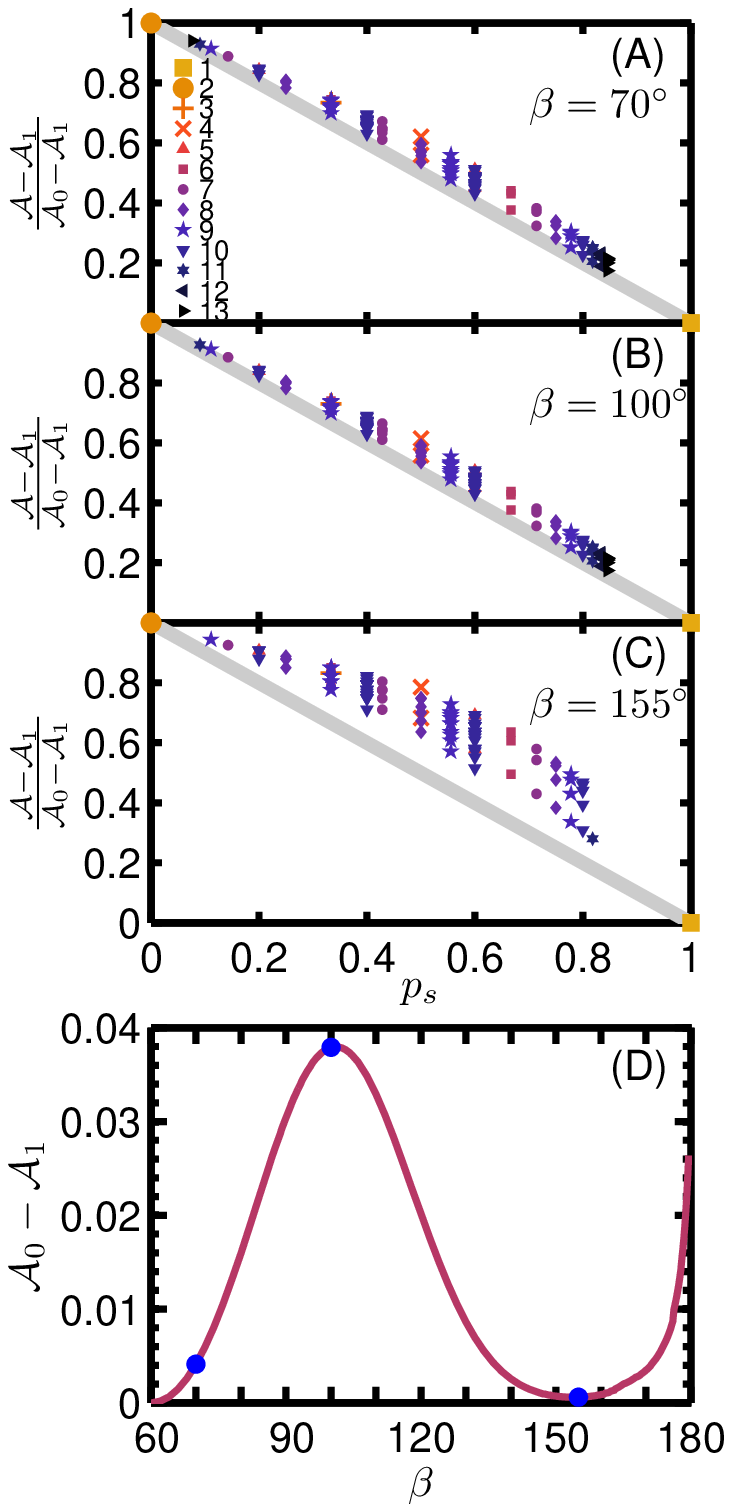}}
\caption{Normalized entropic contribution, Eq.~(\ref{eq:A}) vs fraction
$p_s \equiv P(iii)/[P(iii)+P(iv)]$ of straight-stripe segments for
deformation angles $\beta=70^\circ$(A), $100^\circ$(B), and
$155^\circ$(C). $\mathcal{A}$ is normalized by its extreme values for straight 
stripes $\mathcal{A}_1 \equiv \mathcal{A}(p_s=1)$ and for bent stripes $\mathcal{A}_0 \equiv \mathcal{A}(p_s=0)$. 
The gray lines indicate the prediction of a one-dimensional 
effective Hamiltonian of non-interacting straight and bent segments.
The size of the unit cell corresponding to each ground-state 
configuration is indicated in the legend. We calculated the
free energy for all the 95 distinct unit cells consisting of up to 10
particles. Except for straight ($p_s=1$) and bent ($p_s=0$) stripes,
these have $\frac{1}{9} \le p_s \le \frac{4}{5}$. Additionally, of the
363 unit cells with 11-13 particles, we calculated the free energy for
19 unit cells with extreme values of $p_s$. In numerically evaluating
the integration over the Brillouin zone, we require a relative
numerical accuracy smaller than $10^{-7}$, and therefore for
$\beta=155^\circ$ omitted from the plot 19 unit cells (with 11-13
particles) and for $\beta=100^\circ$ omitted one unit cell (with 13
particles). See SI Appendix for more details on the calculations and 
the results. D) Entropic advantage of straight stripes ($p_s=1$) over
zigzags ($p_s=0$) vs deformation angle. The blue circles indicate the
angles for which results for larger unit cells are given in (A-C).}
\label{fig:free_energy}
\end{figure}

Figure~\ref{fig:free_energy}A-C shows the results obtained by
numerically evaluating Eq.~(\ref{eq:A}) for ground states with unit
cells consisting of up to 13 particles, plotted vs $p_s$ 
(see SI Appendix for more details and Tables S4-S5 for the entire data 
plotted). We find that
the free energy is bounded between the extreme cases of straight
stripes ($p_s=1$, Fig.~\ref{fig:ZZ}A) and bent stripes ($p_s=0$,
Fig.~\ref{fig:ZZ}B). Moreover, the approximate collapse of the results for
all calculated unit cells to a single curve 
provides strong support for the hypothesis that only the ratio
between straight and bent segments is important and not the order in
which they are positioned or the period of the pattern they form.
Finally, the roughly linear dependence of $\mathcal{A}$ on $p_s$ indicates that
the free energy may be approximated as a linear combination of
contributions from the straight and bent segments. Since the ground
state is determined by a one-dimensional sequence of straight and bent
segments, this result implies that the system may be well approximated by an
effective one-dimensional non-interacting Hamiltonian.
For $\beta=155^\circ$ (Fig.~\ref{fig:free_energy}C) we observe substantial deviations from the prediction of such a non-interacting Hamiltonian. It would be interesting to understand whether these deviations result from our limited numerical accuracy in evaluating so small free-energy differences, or from additional considerations that affect the physics of this system at such large deformations.

In our elastic Ising model, the low-temperature free energy of in-plane positional fluctuations is determined by the dimensionless ratio $b\equiv\frac{J\epsilon}{Ka}$ between the magnetoelastic interaction strength to the lattice rigidity. This ratio sets the deformation angle $\beta$ of each isosceles triangle in the ground-state (see Eq.~(\ref{eq:sf}) and Fig.~\ref{fig:model}B). In the corresponding system of buckled colloids, the deformation angle $\beta$ of each isosceles plaquette in the close-packed state (which is equivalent to the Ising model's ground state), is similarly dictated by the ratio of each sphere's diameter to the separation between the confining walls~\cite{Shokef}. We expect the colloidal system to exhibit a similar order-by-disorder effect that will be governed by the deformation angle $\beta$, and therefore we plot in Fig.~\ref{fig:free_energy}D the free-energy difference $\mathcal{A}_0-\mathcal{A}_1$ between straight and bent segments vs the geometrical parameter $\beta$ rather than vs the physical parameter $b$. 

For extremely rigid lattices that hardly deform ($\beta \approx 60^\circ$), straight and zigzagging stripes are almost equivalent in terms of their particle displacements and therefore $\mathcal{A}_0 \approx \mathcal{A}_1$. As $\beta$ increases, straight stripes develop an entropic advantage which comes from the fact that the straight-stripe ground state is the most anisotropic and hence has the most nonuniform distribution of eigen-frequencies, and thus the maximal entropy~\cite{Henley89}. For extremely large deformations ($\beta>100^\circ$) we observe a decrease in $\mathcal{A}_0-\mathcal{A}_1$ which reaches a minimal value at $\beta\approx 155^{\circ}$ and then increases again as $\beta\rightarrow 180^\circ$. In the SI Appendix we show how this non-monotonic behavior results from the numerical structure of the dispersion relations. In particular, we identify in Fig.~S10 the region in reciprocal space which dominates the free-energy difference between straight and bent stripes. The magnitude of the free-energy difference in this region grows monotonically with $\beta$, however the size of this region decreases, which gives rise to the non-monotonic behavior seen in Fig.~\ref{fig:free_energy}D for $60^\circ \le \beta \le 155^\circ$. For $\beta \ge 155^\circ$ the free-energy difference has a different wavevector dependence (see Fig.~S10) which gives rise to the second increase in the total free-energy difference at such large deformations. It would be interesting to theoretically understand the deeper origins and possible implications of this non-monotonic behavior. Note that for large values of $\beta$, the distance to some of the next-nearest neighbors becomes smaller than the distances between nearest neighbors. This introduces further complications beyond the analysis presented in this paper which assumes only nearest-neighbor interactions. Overall, our numerical results show that for all ground-state deformations, straight stripes are preferred entropically, however their entropic advantage is extremely small. We observed the same qualitative behavior when allowing the system's total volume to vary. 

Before concluding we note
that the free energy coefficient $\mathcal{A}$ given in Eq.~(\ref{eq:A}) and
plotted in Fig.~\ref{fig:free_energy} is defined for each specific
realization of the stripes. Randomly zigzagging stripes which mix
straight and bent segments are highly degenerate, and their free energy
possesses also a configurational entropy (or entropy of mixing) which
competes with the vibrational entropic advantage of straight stripes
that we found. However, due to the one-dimensional character of the
ground state, this configurational entropy is sub-extensive in system
size and scales as $\sqrt{N}$. The vibrational free energy of
Eq.~(\ref{eq:A}) enters as an extensive quantity which scales linearly with $N$ in $\mathcal{F}$, and is thus dominant in the thermodynamic limit. As a result, straight
stripes are favored. This may be the reason for the limited ability of
our simulations to reach perfect straight stripes, and for the smaller
values of $P(iii)/P(iv)$ for smaller systems in Fig.~\ref{fig:MC}C. To
test this, we repeated the simulations at parameter values where the
difference in $\mathcal{A}$ between straight stripes and zigzags is smaller and
indeed found a weaker preference for straight stripes.

In summary, our exact microscopic description of this
previously-studied~\cite{Kardar,Bulbul} elastic Ising model reveals a
highly-degenerate, partially-disordered ground state. We find a much
richer behavior since entropy lifts the degeneracy at any positive
temperature. However, large free-energy barriers between the
ground-state configurations induce a glassy phase of zigzagging
stripes. 
Although equilibration is strongly hindered in this glassy phase, the straight-stripe structures obtained in it result from the minute entropic differences in the equilibrium free energies of the various stripe realizations.
The fact that this model is amenable to analytic treatment
makes it appealing as a prototypical model for studying such
order-by-disorder phenomena in a broader context. Moreover, the current
approach to experimentally studying frustration relief by lattice
deformations in antiferromagnets is based on quite indirect
measurements of lattice deformations~\cite{Sagi,Rasch}. Our work makes
direct contact to a colloidal system~\cite{Han} in which frustration
and its relief are governed by similar physical mechanisms, yet it has
the advantage that local deformations can be directly measured in it.
Interestingly, several recent experimental works are focused on measuring the
normal modes of vibration in colloidal systems~\cite{Ghosh2010,Chen2010,Kaya2010,Yunker2011}.
On top of understanding deformable antiferromagnets and mesoscopic
model systems for them, we expect that results obtained for our system
will shed light on questions such as the statistical mechanics of
sphere packings and the physical origins of glassy dynamics.

\begin{acknowledgments}

We thank Ehud Altman, Assa Auerbach, Bulbul Chakraborty, John Chalker,
Kedar Damle, Chris Henley, Randy Kamien, Amit Keren, Joel Lebowitz,
Roderich Moessner, Ido Regev, Per Rikvold, and Peter Yunker for helpful
discussions. This work is supported by NSF MRSEC Grant No. DMR-0520020.

\end{acknowledgments}

\end{article}

\begin{thebibliography}{50} %% Enter the largest bibliography number in the facing curly brackets

\bibitem{PhysicsToday} Moessner R, Ramirez AR (2006) Geometrical
frustration. {\it Phys. Today} 59:24–26.

\bibitem{Wannier} Wannier GH, (1950) Antiferromagnetism. The triangular
Ising net. {\it Phys. Rev.} 79:357–364.

\bibitem{Houtappel} Houtappel RMF (1950) Order-disorder in hexagonal
lattices. {\it Physica} 16:425-455.

\bibitem{Metcalf} Metcalf BD (1974) Ground state spin orderings of the
triangular Ising model with the nearest and next nearest neighbor
interaction. {\it Phys. Lett. A} 46:325-326.

\bibitem{Kardar} Chen ZY, Kardar M (1986) Elastic antiferromagnets on a
triangular lattice. {\it J. Phys. C: Solid State Phys.} 19:6825-6831.

\bibitem{Bulbul} Gu L, Chakraborty B, Garrido PL, Phani M, Lebowitz JL
(1996) Monte Carlo study of a compressible Ising antiferromagnet on a
triangular lattice. {\it Phys. Rev. B} 53:11985-11992.

\bibitem{Terao} Terao K (1996) Effect of lattice distortions upon the
spin configuration of antiferromagnetic YMn$_2$ with C15 structure.
{\it J. Phys. Soc. Jpn.} 65:1413-1417.

\bibitem{Yamashita} Yamashita Y, Ueda K (2000) Spin-driven Jahn-Teller
distortion in a pyrochlore system. {\it Phys. Rev. Lett.} 85:4960-4963.

\bibitem{Tchernyshyov} Tchernyshyov O, Moessner R, Sondhi SL
(2002) Order by distortion and string modes in pyrochlore
antiferromagnets. {\it Phys. Rev. Lett.} 88:067203.

\bibitem{Schilling} Schilling T, Pronk S, Mulder B, Frenkel D (2005)
Monte Carlo study of hard pentagons. {\it Phys. Rev. E} 71:036138.

\bibitem{Villain} Villain J, Bidaux R, Carton JP, Conte R (1980) Order
as an effect of disorder. {\it J. Physique} 41:1263-1272.

\bibitem{Henley87} Henley CL (1987) Ordering by disorder: Ground-state
selection in fcc vector antiferromagnets. {\it J. Appl. Phys.}
61:3962-3964.

\bibitem{Henley89} Henley CL (1989) Ordering due to disorder in a
frustrated vector antiferromagnet. {\it Phys. Rev. Lett.} 62:2056-2059.

\bibitem{Chubukov} Chubukov A (1992) Order from disorder in a kagome
antiferromagnet. {\it Phys. Rev. Lett.} 69:832-835.

\bibitem{Reimers} Reimers JN, Berlinsky AJ (1993) Order by disorder in
the classical Heissenberg kagome antiferromagnet. {\it Phys. Rev. B}
48:9539-9554.

\bibitem{Bergman} Bergman D, Alicea J, Gull E, Trebst S, Balents L
(2007) Order-by-disorder and spiral spin-liquid in frustrated
diamond-lattice antiferromagnets {\it Nature Physics} 3:487-491.

\bibitem{Collins} Collins MF, Petrenko OA (1997) Triangular
antiferromagnets. {\it Can. J. Phys.} 75:605-655.

\bibitem{Maignan} Maignan A, Michel C, Masset AC, Martin C, Raveau B
(2000) Single crystal study of the one dimensional Ca$_3$Co$_2$O$_6$
compound: five stable configurations for the Ising triangular lattice.
{\it Eur. Phys. J. B} 15:657-663.

\bibitem{WangVishwanath} Wang F, Vishwanath A (2008) Spin
phonon induced collinear order and magnetization plateaus in triangular
and Kagome antiferromagnets: applications to CuFeO2. {\it Phys. Rev.
Lett.} 100:077201.

\bibitem{Starykh} Starykh OA, Katsura H, Balents L (2010) Extreme
sensitivity of a frustrated quantum magnet: Cs$_2$CuCl$_4$ {\it Phys.
Rev. B} 82:014421.

\bibitem{Wang} Wang RF et al. (2006) Artificial `spin ice' in a
geometrically frustrated lattice of nanoscale ferromagnetic islands.
{\it Nature} 439:303-306.

\bibitem{Han} Han Y et al. (2008) Geometric frustration in buckled
colloidal monolayers. {\it Nature} 456:898-903.

\bibitem{Chou} Chou T, Nelson DR (1993) Buckling instabilities of a
confined colloid crystal layer. {\it Phys. Rev. E} 48:4611-4621.

\bibitem{Shokef} Shokef Y, Lubensky TC (2009) Stripes, zigzags, and
slow dynamics in buckled hard spheres. {\it Phys. Rev. Lett.}
102:048303.

\bibitem{Schmidt96} Schmidt M, L\"{o}wen H (1996) Freezing between two
and three dimensions. {\it Phys. Rev. Lett.} 76:4552-4555.

\bibitem{Schmidt97} Schmidt M, L\"{o}wen H (1997) Phase diagram of hard
spheres confined between two parallel plates. {\it Phys. Rev. E}
55:7228-7241.

\bibitem{Zangi98} Zangi R, Rice SA (1998) Phase transitions in a
quasi-two-dimensional system. {\it Phys. Rev. E} 58:7529-7544.

\bibitem{Melby} Melby P et al. (2005) The dynamics of thin vibrated
granular layers. {\it J. Phys. Condens. Matter} 17:S2689-S2704.

\bibitem{Stillinger} Stillinger FH, Salsburg ZW (1967) Elasticity of
rigid-disk and -sphere crystals. {\it J. Chem. Phys.} 46:3962-3975.

\bibitem{Alder} Alder BJ, Hoover WG, Young DA (1968) Studies in
molecular dynamics. V. High-density equation of state and entropy for
hard disks and spheres. {\it J. Chem. Phys.} 49:3688-3696.

\bibitem{Rudd} Rudd WG, Salsburg ZW, Yu AP, Stillinger FH (1968) Rigid
disks and spheres at high densities. III {\it J. Chem. Phys.}
49:4857-4863.

\bibitem{Pusey} Pusey PN et al. (1989) Structure of crystals of hard
colloidal spheres {\it Phys. Rev. Lett.} 63:2753-2756.

\bibitem{Woodcock1} Woodcock LV (1997) Entropy difference between the
face-centred cubic and hexagonal close-packed crystal structures. {\it
Nature} 385:141-143.

\bibitem{Bolhuis} Bolhuis PG, Frenkel D, Mau SC, Huse DA (1997) Entroy
difference between crystal phases. {\it Nature} 388:235-236.

\bibitem{Woodcock2} Woodcock LV (1997) Entropy difference between
crystal phases - Reply. {\it Nature} 388:236-237.

\bibitem{Mau} Mau SC, Huse DA (1999) Stacking entropy of hard-sphere
crystals. {\it Phys. Rev. E} 59:4396-4401.

\bibitem{Radin} Radin C, Sadun L (2005) Structure of the hard sphere
solid. {\it Phys. Rev. Lett.} 94:015502.

\bibitem{Koch} Koch H, Radin C, Sadun L (2005) Most stable structure
for hard spheres. {\it Phys. Rev. E} 72:016708.

\bibitem{Ogawa} Ogawa T (1983) A maze-like pattern in a monodispersive
latex system and the frustration problem. {\it J. Phys. Soc. Jpn.
Suppl.} 52:167-170.

\bibitem{jen} Lynch JM et al. (2010) Glassy dynamics of geometrically
frustrated colloidal system. Presented at the APS March Meeting,
Portland OR, March 2010.
http://meetings.aps.org/link/BAPS.2010.MAR.D12.11

\bibitem{Tome} Tome T, de Oliveira MJ (1990) Dynamic phase transition in the kinetic Ising model under a time-dependent oscillating field. {\it Phys. Rev. A} 41:4251-4254.

\bibitem{Sides} Sides SW, Rikvold PA, Novotny MA (1998) Stochastic hysteresis and resonance in a kinetic Ising system. {\it Phys. Rev. E} 57:6512-6533.

\bibitem{Korniss} Korniss G, White CJ, Rikvold PA, Novotny MA (2000) Dynamic phase transition, universality, and finite-size scaling in the two-dimensional kinetic Ising model in an oscillating field. {\it Phys. Rev. E} 63:016120.

\bibitem{Robb} Robb DT et al. (2008) Evidence for a dynamic phase transition in [Co/Pt]$_3$ magnetic multilayers. {\it Phys. Rev. B} 78:134422.

\bibitem{Sagi} Sagi E, Ofer O, Keren A, Gardner JS (2005) Quest for
frustration driven distortion in Y$_2$Mo$_2$O$_7$ {\it Phys. Rev.
Lett.} 94:237202.

\bibitem{Rasch} Rasch JCE et al. (2009) Magnetoelastic coupling in the
triangular lattice antiferromagnet CuCrS$_2$. {\it Phys. Rev. B}
80:104431.

\bibitem{Ghosh2010} Ghosh A, Chikkadi VK, Schall P, Kurchan J, Bonn D (2010) Density of States of colloidal glasses {\it Phys. Rev. Lett.} 104:248305.

\bibitem{Chen2010} Chen K, Ellenbroek WG, Zhang Z, Chen DTN, Yunker PJ, Henkes S, Brito C, Dauchot O, van Saarloos W, Liu AJ, Yodh AG (2010) Low-frequency vibrations of soft colloidal glasses {\it Phys. Rev. Lett.} 105:025501.

\bibitem{Kaya2010} Kaya D, Green NL, Maloney CE, Islam MF (2010) Normal modes and density of states of disordered colloidal solids {\it Science} 329:656.

\bibitem{Yunker2011} Yunker PJ, Chen K, Zhang Z, Yodh AG (2011) Phonon spectra, nearest neighbors, and mechanical stability of disordered colloidal clusters with attractive interactions
arXiv:1103.3535

\end{thebibliography}
\end{document}